\documentclass[twocolumn]{revtex4}
\usepackage[dvips]{graphicx}
\usepackage{float, color,array,amsfonts}
\usepackage{float}

\begin{document}

\title{Stripe order and diode effect in two-dimensional Rashba superconductors}

\author{Kazushi Aoyama}

\affiliation{Department of Earth and Space Science, Graduate School of Science, Osaka University, Osaka 560-0043, Japan}

\begin{abstract}
In two-dimensional superconductors with a Rashba-type spin-orbit coupling, it is known that an in-plane magnetic field can induce a helical superconducting (SC) state with a phase modulation $e^{i {\bf q}\cdot {\bf r}}$. Here, we theoretically investigate the stability of a stripe order, a weight-biased superposition state composed of $+{\bf q}$ and $-{\bf q}$ modes taking the form of $\Delta_+ e^{i{\bf q}\cdot{\bf r}}+\Delta_- e^{-i{\bf q}\cdot{\bf r}}$ with $|\Delta_+|\neq|\Delta_-|\neq 0$, assuming that the spin-singlet pairing channel is dominant. Based on the Ginzburg-Landau theory, we show that for both $s$-wave and $d$-wave pairing symmetries, the stripe order can appear in the high-field and low-temperature region inside the helical phase and that the transition between the helical and stripe phases is of second order. It is noteworthy that for the $d$-wave pairing, the stability region of the stripe phase shrinks when the in-plane field is rotated from the nodal direction to the anti-nodal direction. It is also found that the nonreciprocity of the critical current, the so-called SC diode effect, emerges not only in the helical phase but also in the stripe phase, with no clear nonreciprocity anomaly at the helical-stripe transition due to its second-order nature. 
\end{abstract}

\maketitle

\section{Introduction}
In superconductors and Fermionic superfluids, Cooper pairs with their center-of-mass momenta ${\bf q}$'s are usually condensed into a uniform state of ${\bf q}=0$, as ${\bf q}\neq 0$ modes and associated spatial variations generally require additional energy cost. When the Cooper pair is subject to pair-breaking effects due to an external magnetic field \cite{FF, LO, FFLO_Matsuda_review_07} and surface scatterings \cite{Vorontsov_SF, Vorontsov_SC, Hachiya, Aoyama_cylinder, Aoyama_film, Wiman_film}, it can happen that the Cooper-pair condensate chooses to have a ${\bf q}\neq 0$ rather than to be uniformly suppressed. A prime example of such a ${\bf q}\neq 0$ superconducting (SC) state is the Fulde-Ferrell-Larkin-Ovchinnikov (FFLO) state induced by the Pauli-paramagnetic pair-breaking effect \cite{FF, LO, FFLO_Matsuda_review_07} where the SC gap function exhibits a phase modulation $e^{i {\bf q} \cdot {\bf r}}$ (FF state) or an amplitude modulation $\cos({\bf q} \cdot {\bf r})$ (LO state). The latter can be viewed as a superposition of $\pm {\bf q}$ modes of equal weight. In this work, we theoretically investigate the stability of a weight-biased superposition state called a stripe order \cite{Stripe_Agterberg_prb_07, 2layer-Stripe_Yoshida_jpsj_13} in two-dimensional non-centrosymmetric superconductors possessing a Rashba-type spin-orbit coupling (RSOC) where the degeneracy between ${\bf q}$ and $-{\bf q}$ is lifted by a combined effect of the RSOC and an in-plane Zeeman field \cite{Kaur,Springer} and resultantly, a supercurrent becomes nonreciprocal \cite{diode_Wakatsuki_prl_18, diode_Wakatsuki_sadv_17}, giving rise to a SC diode effect \cite{diode_Ando_nature_20, diode_Daido_prl_22, diode_Yuan_pnas_22, diode_He_njp_22, diode_Daido_prb_22}.

The RSOC of the form $\alpha_{\rm R} \left( {\bf k} \times \hat{z} \right)\cdot {\mbox {\boldmath $\sigma$}}$ is generic to non-centrosymmetric systems lacking mirror symmetry with respect to a crystalline plane such as the heavy-fermion bulk superconductors CePt$_3$Si \cite{CePt3Si} and CeTSi$_3$ (T=Rh, Ir) \cite{CeRhSi3, CeIrSi3} and two-dimensional superconductors realized on artificial superlattices \cite{superlattice_Goh_prl_12, superlattice_Shimozawa_prl_14, superlattice_Naritsuka_prb_17}, at interfaces between two different materials \cite{interface-SC, tunable-SC, tunable-RSO}, and in perpendicular electrical gate fields \cite{gateE-SC_Ueno_natmat_08, gateE-SC_Ueno_prb_14}. In a magnetic field ${\bf H}$ parallel to the mirror plane, a field-induced in-plane spin polarization is connected to supercurrents via the RSOC \cite{SC_jM, SC_Bj, Dimitrova,Samokhin,Kaur,Fujimoto,Helical_AS_12,Helical_ASS_14}. Such a magneto-electric effect can induce a so-called helical SC state with a phase modulation $e^{i {\bf q}\cdot {\bf r}}$ \cite{Kaur}, where in contrast to the conventional FF state, the direction of ${\bf q}$ is fixed to be parallel to $\alpha_{\rm R} \left( {\bf H}\times \hat{z}\right)$ due to the RSOC and thus, ${\bf q}$ and $-{\bf q}$ are not equivalent. Noting that the phase gradient yields a SC current, it turns out that the nonequivalence between ${\bf q}$ and $-{\bf q}$ should be reflected as a nonreciprocity between the supercurrents flowing in the ${\bf q}$ and $-{\bf q}$ directions. Actually, a nonreciprocal SC transport of this kind has been observed as the SC diode effect \cite{diode_Ando_nature_20} which can be understood as the nonreciprocity of the SC critical current \cite{diode_Daido_prl_22, diode_Yuan_pnas_22, diode_He_njp_22, diode_Daido_prb_22}. Here, we emphasize again that the key ingredient for the helical phase and the associated SC diode effect is the RSOC. 

In the case without the RSOC, on the other hand, it is well known that the Zeeman field or the Pauli-paramagnetic pair-breaking effect leads to the occurrence of the amplitude modulated LO state in a high-field and low-temperature region \cite{FFLO_Matsuda_review_07}. Since as mentioned above, the RSOC favors the FF-like helical state, how robust the LO state is against the RSOC would be an interesting question. This issue has already been discussed in three-dimensional Rashba superconductors with a dominant spin-singlet $s$-wave pairing interaction \cite{Stripe_Agterberg_prb_07}, although in three dimensions, the orbital pair-breaking effect, which is not incorporated in Ref. \cite{Stripe_Agterberg_prb_07}, is non-negligible \cite{Ikeda_0}. It has been reported that an intermediate state between the FF and LO states taking the form of $\Delta_+ e^{i{\bf q}\cdot  {\bf r}}+ \Delta_- e^{-i{\bf q}\cdot {\bf r}}$ with $|\Delta_+|\neq |\Delta_-|\neq 0$ can be stabilized \cite{Stripe_Agterberg_prb_07}. Since this intermediate state breaks the translational symmetry similarly to the LO state of $|\Delta_+| = |\Delta_-|$, it is called the stripe order, being distinguished from the conventional LO state. In contrast to the helical phase where only one modulation vector ${\bf q}$ determines the preferred flow-direction of the supercurrent, the stripe phase involves both ${\bf q}$ and $-{\bf q}$, so that how the SC diode effect looks like in the stripe phase would be an interesting question. Furthermore, when we consider a $d$-wave pairing instead of the isotropic $s$-wave one, we notice that another fundamental question arises.    

\begin{figure*}[t]
\centering
\includegraphics[width=2\columnwidth]{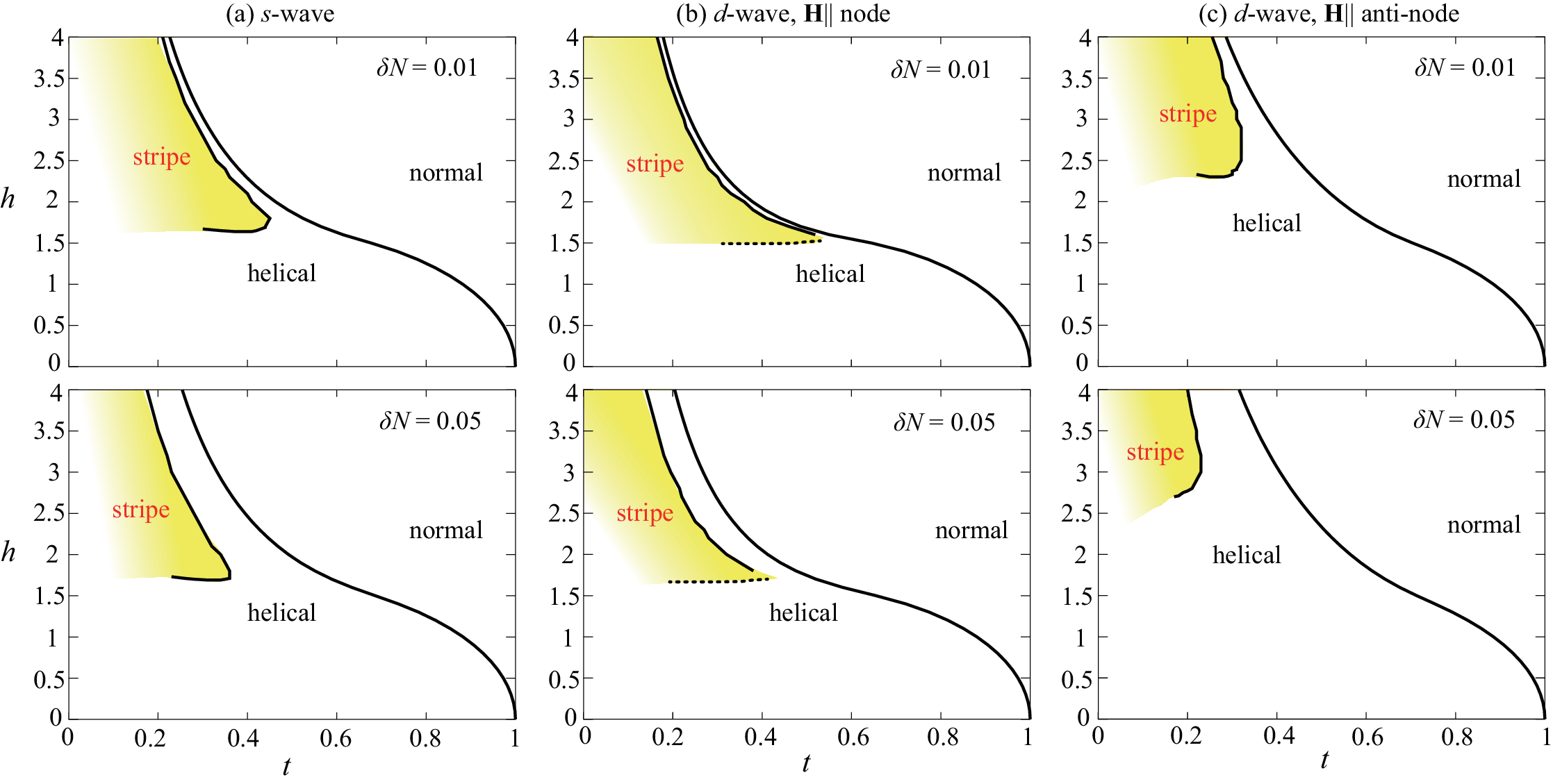}
\caption{Temperature and magnetic-field phase diagrams of the two-dimensional Rashba superconductors with spin-singlet pairing interactions whose orbital symmetries are (a) $s$-wave, and (b) and (c) $d$-wave of $d_{x^2-y^2}$ type. In (b) [(c)], the in-plane field is applied parallel to the nodal [110] direction (the anti-nodal [100] direction). In the upper and lower panels, $\delta N$'s measuring the strength of the RSOC are $0.01$ and $0.05$, respectively. The solid lines denote the second order transitions. In (b), the lower boundary of the stripe phase cannot be determined in the present GL approach (for details, see Appendix A), and thus, the dashed line is just a guide for eyes. \label{fig:HT}}
\end{figure*}

In $d$-wave superconductors, there exists a spatial anisotropy originating from of the SC gap nodes. In the $d_{x^2-y^2}$ case {\it without} the RSOC, it has been theoretically shown that for a cylindrical Fermi surface, the LO state with ${\bf q}$ parallel to the nodal [110] direction is stabilized in the high-field and low-temperature region and that at further high fields, the FF state with ${\bf q}$ parallel to the anti-nodal [100] direction becomes more stable \cite{FFLO_d_Maki_physicaB_02, FFLO_Vorontsov_prb_05}. We note that the direction of the applied field has nothing to do with the ${\bf q}$ direction as the spin and orbital sectors are completely decoupled in the absence of a spin-orbit coupling. Then, the question is what happens when the RSOC is introduced. Since the RSOC favors the phase-modulated helical state with ${\bf q}\parallel \alpha_{\rm R} \left( {\bf H}\times \hat{z}\right)$, the field direction relative to the nodal direction should be important for the stability of the LO-like stipe phase involving both ${\bf q}$ and $-{\bf q}$. To our knowledge, this issue has not yet been discussed. In addition, a recent experiment on the globally-noncentrosymmetric tricolor superlattice of YbRhIn$_5$/CeCoIn$_5$/YbCoIn$_5$ \cite{superlattice_Naritsuka_prb_17} containing CeCoIn$_5$ whose pairing symmetry in the bulk is $d_{x^2-y^2}$ \cite{VLFF, Hiasa, Magreso_Stock_prl_08, angledep_Vorontsov, d-wave, node-imaging_Allan_natphys_13, QPI_Zhou_natphys_13} has shown that a nonreciprocal SC transport exhibits an in-plane anisotropy when the in-plane field is rotated from the nodal ([110]) direction to the anti-nodal ([100]) direction \cite{superlattice_Matsuda_private}. In view of such theoretical and experimental situations, we theoretically investigate the stripe instability in the two-dimensional Rashba superconductors for the $d$-wave pairing and the $s$-wave one as well for reference, and examine the SC diode effect in the stripe phase.

Based on the Ginzburg-Landau (GL) analysis, we will show that for ${\bf H}$ parallel to the nodal [110] direction, the stripe phase can be stabilized in a relatively wide range of the high-field and low-temperature region, while not for ${\bf H}$ parallel to the anti-nodal [100] direction [see Figs. \ref{fig:HT} (b) and (c)]. In both cases, with increasing $\delta N$ measuring the strength of the RSOC (see Sec. II), the stripe phase gets unstable against the helical phase. It is also found that the nonreciprocity of the critical SC current, i.e., the intrinsic SC diode effect \cite{diode_Daido_prl_22, diode_Yuan_pnas_22, diode_He_njp_22, diode_Daido_prb_22}, emerges in the stripe phase as well as in the helical phase and that the second order transition between the two phases does not leave a footprint in the temperature dependence of the critical-current nonreciprocity. 

This paper is organized as follows: In Sec. II, we derive the GL free energy functional from the microscopic BCS Hamiltonian with the RSOC and explain how to examine the stability of the stripe order and the nonreciplocity of the critical current. Results on the former and latter issues are discussed in Secs. III and IV, respectively. We end the paper with summary and discussions in Sec. V.

\section{Theoretical Framework}
In this section, we derive the GL free energy functional from the microscopic Hamiltonian, and explain how to examine the stability of the stripe order and the nonreciplocity of the critical current.
\subsection{GL free energy functional}
In this work, we consider the two-dimensional Rashba superconductor with the spin-singlet pairing interaction. In the in-plane magnetic field ${\bf H}$, the microscopic BCS Hamiltonian can be written as
\begin{eqnarray}\label{eq:Hamiltonian}
&& {\cal H} = \sum_{{\bf k},s,s'} K_{ss'} ({\bf k}) \hat{c}_{{\bf k},s}^{\dag} \hat{c}_{{\bf k},s'} - U  \sum_{\bf q} \, \hat{B}^\dagger({\bf q}) \hat{B}({\bf q}),\\
&& K_{ss'}({\bf k}) = \varepsilon_{\bf k} \delta_{ss'}+ {\mbox {\boldmath $\sigma$}}_{ss'}\cdot \big( \alpha \, {\bf g}_{\bf k} - \mu {\bf H}\big), \nonumber\\
&& \hat{B}({\bf q}) = \frac{1}{2} \sum_{{\bf k},s,s'} (-i \, \sigma_y)_{ss'} \, w_{\bf k} \hat{c}_{-{\bf k}+\frac{\bf q}{2},s} \hat{c}_{{\bf k}+\frac{\bf q}{2},s'}, \nonumber
\end{eqnarray}
where $\hat{c}^\dagger_{{\bf k} ,s}$ and $\hat{c}_{{\bf k} ,s}$ are the creation and annihilation operators of a quasiparticle with momentum ${\bf k}$ and spin $s$, respectively, and ${\mbox {\boldmath $\sigma$}}=(\sigma_x,\sigma_y,\sigma_z)$ with Pauli matrices $\sigma_i$.
The single-particle spectrum is described by $ K_{ss'} $ which includes the RSOC with strength $\alpha_{\rm R}>0$ and the unit vector ${\bf g}_{\bf k} = \hat{k} \times \hat{z} $ ($ \hat{k} = {\bf k}/k_F$) and the Zeeman field $\mu {\bf H}$. The kinetic energy $ \varepsilon_{\bf k} $ is measured from the Fermi energy $ E_F $. In Eq. (\ref{eq:Hamiltonian}), the last term represents the spin-singlet pairing interaction with its orbital symmetry being described by $w_{\bf k}$ which takes the form of $w_{\bf k}=1$ and $w_{\bf k}=\sqrt{2} \, \cos(2\phi_{\bf k})$ in the $s$-wave and $d_{x^2-y^2}$ cases, respectively, where $\phi_{\bf k}$ is an azimuthal angle in the $k_x$-$k_y$ plane. 
In this work, the Fermi surface is assumed to be isotropic, or equivalently, cylindrical, so that in the absence of the RSOC, the anisotropy of the system enters only via $w_{\bf k}$. 

By introducing the unitary transformation to diagonalize the single-particle Hamiltonian $\hat{c}_{{\bf k},s}=\frac{1}{\sqrt{2}}(-i \sin\phi'+\sigma_z \cos\phi'+\sigma_x) _{ss'} \hat{b}_{{\bf k},s'}$ with $\phi'=\cos^{-1}\frac{( \alpha_{\rm R} \, {\bf g}_{\bf k} - \mu {\bf H})_x}{|\alpha_{\rm R} \, {\bf g}_{\bf k} - \mu {\bf H}|}$
and the mean field $\Delta({\bf q})=U\langle \hat{B}({\bf q}) \rangle$, Eq. (\ref{eq:Hamiltonian}) can be rewritten in the mean-field approximation as
\begin{eqnarray}
{\cal H}_{\rm MF} &=& \frac{1}{U}\sum_{\bf q} |\Delta({\bf q})|^2 + \sum_{a=1,2}\Big( \sum_{\bf k}\xi_{\bf k}^a \,\hat{b}_{{\bf k},a}^\dagger \hat{b}_{{\bf k},a} \nonumber\\
&+& \frac{1}{2}\sum_{{\bf k},{\bf q}}\Delta({\bf q}) e^{-i(-1)^{a} (\phi_{\bf k}+\frac{\pi}{2})}\, w_{\bf k}^\ast \, \hat{b}_{{\bf k}+\frac{\bf q}{2},a}^\dagger \hat{b}_{-{\bf k}+\frac{\bf q}{2},a}^\dagger \nonumber\\
&+& \frac{1}{2}\sum_{{\bf k},{\bf q}}\Delta^\ast({\bf q}) e^{i(-1)^{a} (\phi_{\bf k}+\frac{\pi}{2})}\, w_{\bf k} \, \hat{b}_{-{\bf k}+\frac{\bf q}{2},a} \hat{b}_{{\bf k}+\frac{\bf q}{2},a} \Big), \nonumber
\end{eqnarray}
where $a=1,2$ denotes the two Fermi surfaces split by the RSOC. Here, we have used the assumptions $\mu H \ll \alpha_{\rm R} \ll E_F$ and $|{\bf q}|\ll k_F$ the former of which yields 
\begin{eqnarray}
\xi_{\bf k}^a &=& \varepsilon_{\bf k} - (-1)^{a}\left|\alpha_{\rm R} \, {\bf g}_{\bf k} - \mu {\bf H}\right| \nonumber\\
&\simeq& \varepsilon_{\bf k} - (-1)^{a} \big[ \alpha_{\rm R} - (\mu{\bf H}\times \hat{k})_z \big] 
\end{eqnarray}
and ${\bf v}^a_{\bf k}= v_F \big[ \hat{k} -(-1)^a \frac{\mu H}{v_F k_F}(\hat{H}\times \hat{z})\big] \simeq {\bf v}_F$ for the quasiparticle velocity.
Noting that due to the RSOC-induced energy shift, the density of states for the Fermi surface 2, $N_2$, is greater than that for the Fermi surface 1, $N_1$, it turns out that $\delta N = \frac{N_2-N_1}{N_2+N_1}>0$ indirectly measures the strength of the RSOC. The averaged quantity $\overline{N}_0=(N_1+N_2)/2$, on the other hand, enters as a prefactor in the free energy, so that it does not affect the SC instability. 

By expanding the free energy ${\cal F}=- \frac{1}{\beta} \ln {\rm tr} \big[ \exp(-\beta {\cal H}_{\rm MF}) \big]$ up to the fourth order in $\Delta$, we obtain the GL free energy density ${\cal F}_{\rm GL} /V = \overline{N}_0\big[ f_{\rm GL}^{(2)} + f_{\rm GL}^{(4)} \big]$ as
\begin{eqnarray}
f_{\rm GL}^{(2)} &=& \frac{1}{V} \int_{\bf r} \Delta^\ast({\bf r}) \, a^{(2)}({\mbox {\boldmath $\Pi$}}) \, \Delta({\bf r}), \nonumber\\
f_{\rm GL}^{(4)} &=& \frac{1}{V} \int_{\bf r} a^{(4)}({\mbox {\boldmath $\Pi$}}^\dagger_1, {\mbox {\boldmath $\Pi$}}_2, {\mbox {\boldmath $\Pi$}}^\dagger_3) \, \Delta^\ast({\bf s}_1) \Delta({\bf s}_2) \Delta^\ast({\bf s}_3) \Delta({\bf s}_4) \big|_{{\bf s}_i= {\bf r}} \nonumber
\end{eqnarray}
with 
\begin{widetext}
\begin{eqnarray}\label{eq:GL_original}
&& a^{(2)}({\mbox {\boldmath $\Pi$}}) = \ln\frac{T}{T_c}+ \int_0^\infty d\rho  \Big\langle f_{\cos}(\rho,0)-|w_{\bf k}|^2 f_{\cos}(\rho,{\bf H}) \cos({\bf v}_F\cdot {\mbox {\boldmath $\Pi$}} \rho) +\delta N |w_{\bf k}|^2 f_{\sin}(\rho,{\bf H}) \sin({\bf v}_F\cdot {\mbox {\boldmath $\Pi$}} \rho) \Big\rangle_{\rm FS}, \nonumber\\
&& a^{(4)}({\mbox {\boldmath $\Pi$}}^\dagger_1, {\mbox {\boldmath $\Pi$}}_2, {\mbox {\boldmath $\Pi$}}^\dagger_3) =  \frac{1}{2} \int_0^\infty \prod_{i=1}^3 d\rho_i \sum_{k=A,B}\Big\langle  |w_{\bf k}|^4 f_{\cos}(\rho_1+\rho_2+\rho_3,{\bf H}) \cos\big( {\bf v}_F\cdot {\mbox {\boldmath $\Pi$}}^\dagger_1 \, \eta_{k,1} + {\bf v}_F\cdot {\mbox {\boldmath $\Pi$}}_2 \, \eta_{k,2} + {\bf v}_F\cdot {\mbox {\boldmath $\Pi$}}^\dagger_3 \, \eta_{k,3} \big) \nonumber\\
&& \qquad\qquad\qquad\qquad\qquad\qquad - \delta N   |w_{\bf k}|^4  f_{\sin}(\rho_1+\rho_2+\rho_3,{\bf H}) \sin\big( {\bf v}_F\cdot {\mbox {\boldmath $\Pi$}}^\dagger_1 \, \eta_{k,1} + {\bf v}_F\cdot {\mbox {\boldmath $\Pi$}}_2 \, \eta_{k,2} + {\bf v}_F\cdot {\mbox {\boldmath $\Pi$}}^\dagger_3 \, \eta_{k,3} \big) \Big\rangle_{\rm FS} ,
\end{eqnarray}
\end{widetext}
\begin{eqnarray}
&& f_{\cos}(X,{\bf H})=\frac{2\pi T }{\sinh(2\pi T X)}\cos\big(2 (\mu{\bf H}\times \hat{k})_z \, X \big), \nonumber\\
&& f_{\sin}(X,{\bf H})=\frac{2\pi T }{\sinh(2\pi T X)}\sin\big( 2 (\mu{\bf H}\times \hat{k})_z \, X \big), \nonumber\\
&& {\mbox {\boldmath $\eta$}}_A=(\rho_1,\rho_2,\rho_3), \quad {\mbox {\boldmath $\eta$}}_B=(\rho_1+\rho_2, -\rho_2,\rho_2+\rho_3), \nonumber
\end{eqnarray} 
where $\langle {\cal O} \rangle_{\rm FS}=\frac{1}{2\pi}\int_0^{2\pi}d\phi_{\bf k} {\cal O}$ represents the angle average on the Fermi surface. For later convenience, the replacement ${\bf q} \rightarrow {\mbox {\boldmath $\Pi$}}=-i\nabla + 2|e|{\bf A}$ has been used to formally take the gauge field ${\bf A}$ into account, and ${\mbox {\boldmath $\Pi$}}_i$ in Eq. (\ref{eq:GL_original}) acts only on ${\bf s}_i$. For more details, see Ref. \cite{Ikeda_0, Ikeda, AFMSC_KA_prb_11} where the essentially the same derivation method has been used in the different context of SC vortex lattices. In contrast to the usual GL theory where the free energy is also expanded with respect to ${\mbox {\boldmath $\Pi$}}$, here, we have incorporated all the higher-order contributions without using the expansion in ${\mbox {\boldmath $\Pi$}}$, as a higher-order contribution is already known to be important for the SC diode effect \cite{diode_Daido_prl_22, diode_Yuan_pnas_22, diode_He_njp_22}. We note that when $a^{(2)}({\mbox {\boldmath $\Pi$}})$ is expanded with respect to ${\mbox {\boldmath $\Pi$}}$ and $\mu {\bf H}$, the leading order contribution proportional to $\delta N$ turns out to give the Lifshitz invariant \cite{Springer, Stripe_Dimitrova_prb_07}, the origin of the magneto-electric effect in the Rashba superconductors.  

From ${\cal F}_{\rm GL}$, one can calculate the SC current ${\bf j}({\bf r})=-\frac{\delta {\cal F}_{\rm GL}}{\delta {\bf A}}$ as
\begin{eqnarray}\label{eq:current}
&& {\bf j}({\bf r}) = -2|e| \overline{N}_0 \Delta^\ast({\bf r}) {\bf K} ({\mbox {\boldmath $\Pi$}}) \Delta({\bf r}), \nonumber\\
&& {\bf K}({\mbox {\boldmath $\Pi$}}) = \int_0^\infty d\rho\rho \Big\langle {\bf v}_F |w_{\bf k}|^2 \Big\{ f_{\cos}(\rho,{\bf H})\sin({\bf v}_F\cdot {\mbox {\boldmath $\Pi$}} \rho) \nonumber\\
&& \qquad\qquad + \delta N f_{\sin}(\rho,{\bf H})\cos({\bf v}_F\cdot {\mbox {\boldmath $\Pi$}} \rho) \Big\} \Big\rangle_{\rm FS} .
\end{eqnarray}

\subsection{Stability of the stripe order}
In this work, we examine the stability of the stripe order taking the form of
\begin{equation}\label{eq:gap_stripe}
\Delta({\bf r})=\Delta_+ e^{i {\bf q}\cdot {\bf r}} +\Delta_- e^{-i{\bf q} \cdot {\bf r}} .
\end{equation}
For $\Delta_-=0$, it becomes the helical SC state with a phase modulation similar to the FF state, whereas for $|\Delta_+|=|\Delta_-|$, it becomes the LO state with an amplitude modulation. In the former and latter SC states, the time-reversal and translational symmetries are broken, respectively. The stripe order of our interest is described as a weight-biased superposition, i.e., $|\Delta_+| \neq |\Delta_-| \neq 0$, and thus, both the time-reversal and translational symmetries are broken in the stripe phase.

By substituting Eq. (\ref{eq:gap_stripe}) into ${\cal F}_{\rm GL}$, we obtain the free energy density $f_{\rm GL}=f^{(2)}_{\rm GL}+f^{(4)}_{\rm GL}$ as
\begin{eqnarray}\label{eq:fGL_stripe}
f_{\rm GL} &=& \alpha({\bf q})|\Delta_+|^2 + \beta({\bf q})|\Delta_+|^4  \\
&+& \alpha(-{\bf q})|\Delta_-|^2 + \beta(-{\bf q})|\Delta_-|^4  + \gamma ({\bf q}, -{\bf q}) |\Delta_+|^2 |\Delta_-|^2 \nonumber
\end{eqnarray}
with the coefficients 
\begin{eqnarray}\label{eq:fGL_coefficient}
&& \alpha({\bf q}) = a^{(2)}({\bf q}), \quad \beta({\bf q}) = a^{(4)}({\bf q},{\bf q},{\bf q}), \\
&& \gamma({\bf q}, -{\bf q})  = a^{(4)}({\bf q},{\bf q},-{\bf q})+ a^{(4)}({\bf q},-{\bf q},-{\bf q}) \nonumber\\
&& \qquad \qquad \,  + \, a^{(4)}(-{\bf q},{\bf q},{\bf q})+a^{(4)}(-{\bf q},-{\bf q},{\bf q})  .\nonumber 
\end{eqnarray}
In Eq. (\ref{eq:fGL_stripe}), the first line describes the free energy for the helical state of $\Delta_-=0$, $f_{\rm GL, H}$, from which an optimal modulation ${\bf q}={\bf Q}$ is determined such that $f_{\rm GL, H}$ is minimized and then, the gap amplitude is given by $|\Delta_+|^2 = - \frac{1}{2} \frac{\alpha({\bf Q})}{\beta({\bf Q})}$. The onset of the superconductivity or the $H_{c2}$ curve is determined by the condition $\alpha({\bf Q})=0$. 
In the presence of the RSOC ($\delta N \neq 0$), $\alpha(-{\bf Q})$ is always positive at $\alpha({\bf Q})=0$, so that the stripe order involving both ${\bf Q}$ and $-{\bf Q}$ is not realized at least just below the $H_{c2}$ curve. The stripe phase, if it exists, is always preempted by the helical phase. 

We examine the stability of the stripe state against the helical state in essentially the same procedure as that used in the three-dimensional $s$-wave case \cite{Stripe_Agterberg_prb_07}. First, we determine the modulation ${\bf Q}$ and the SC gap $|\Delta_+|$ for the helical state. Noting that the second line in Eq. (\ref{eq:fGL_stripe}) can be rewritten as $\alpha^{\rm eff}(-{\bf q})|\Delta_-|^2 + \beta(-{\bf q})|\Delta_-|^4 $ with
\begin{equation}\label{eq:fGL_stripe_re}
\alpha^{\rm eff}(-{\bf q}) = \alpha(-{\bf q})+\gamma({\bf q},-{\bf q})|\Delta_+|^2,
\end{equation}
it turns out that $\alpha^{\rm eff}(-{\bf q})$ effectively works as a net quadratic term for $\Delta_-$ and the condition $\alpha^{\rm eff}(-{\bf Q})=0$ determines the onset of the stripe order with $\Delta_- \neq 0$. Whether the transition into the stripe phase is of second order or not can be identified from the sign of the $|\Delta_-|^4$ term, $\beta(-{\bf Q})$. 

Inside the stripe phase, the modulation ${\bf q}$ does not have to be the same as ${\bf Q}$ determined for the helical state. Nevertheless, as will be shown in Sec. III, the transition from the helical phase into the stripe phase is of second order, so that the modulations in the two phases would not differ so much near the transition. Thus, in this work, the modulation ${\bf Q}$ obtained for the helical state tentatively assumed over the phase diagram is used as the modulation for the stripe phase.  

\subsection{Nonreciprocity of the critical current}
The critical supercurrents in the stripe and helical phases can be obtained in the same manner as that widely used elsewhere \cite{diode_Daido_prl_22, diode_Yuan_pnas_22, diode_He_njp_22, Thinkham, LP_KA_prb}. We first extend Eq. (\ref{eq:gap_stripe}) into the current-flowing SC state of the form 
\begin{equation}\label{eq:gap_stripe_current}
\Delta({\bf r})=\big[ \Delta_+ e^{i{\bf Q}\cdot{\bf r}}+\Delta_- e^{-i{\bf Q}\cdot {\bf r}} \big] e^{i{\bf q}_{\rm ex}\cdot{\bf r}},
\end{equation} 
where an external current is applied along the ${\bf q}_{\rm ex}$ direction. 
By substituting Eq. (\ref{eq:gap_stripe_current}) into Eq. (\ref{eq:current}), we obtain the SC current averaged over the space ${\bf j} = \frac{1}{V}\int_{\bf r} {\bf j}({\bf r})$ as 
\begin{equation}\label{eq:current_fin}
{\bf j}  = -2|e|\overline{N}_0 \Big[ {\bf K}({\bf Q}_+) |\Delta_+|^2 + {\bf K}(-{\bf Q}_-) |\Delta_-|^2 \Big] 
\end{equation}
with ${\bf Q}_\pm ={\bf Q}\pm{\bf q}_{\rm ex}$. Note that $|\Delta_+|$ and $|\Delta_-|$ are also dependent on ${\bf Q}_\pm$.
In the helical phase, $|\Delta_+|^2 = - \frac{1}{2} \frac{\alpha({\bf Q}_+)}{\beta({\bf Q}_+)}$ and $|\Delta_-|^2=0$, whereas in the stripe phase, $\Delta_+$ and $\Delta_-$ are determined from the coupled GL equations $\frac{\partial f_{\rm GL}}{\partial \Delta_+}=0$ and $\frac{\partial f_{\rm GL}}{\partial \Delta_-}=0$ as
\begin{eqnarray}\label{eq:GL_sol}
|\Delta_+|^2 &=& \frac{-1}{D} \big[ 2\alpha({\bf Q}_+)\beta(-{\bf Q}_-)-\alpha(-{\bf Q}_-)\gamma({\bf Q}_+,-{\bf Q}_-) \big], \nonumber\\
|\Delta_-|^2 &=& \frac{-1}{D} \big[ 2\alpha(-{\bf Q}_-)\beta({\bf Q}_+)-\alpha({\bf Q}_+)\gamma({\bf Q}_+,-{\bf Q}_-) \big], \nonumber\\
D &=& 4 \beta({\bf Q}_+) \beta(-{\bf Q}_-)-\big[ \gamma({\bf Q}_+, -{\bf Q}_-)\big]^2 .
\end{eqnarray}
In this work, we consider the situation where the external current is applied along the $\pm{\bf Q}$ directions perpendicular to the external field ${\bf H}$, and introduce the notations ${\bf q}_{\rm ex}=q_{\rm ex} \, \hat{Q}$ and ${\bf j} = j \, \hat{Q}$. Then, the maximum value of the supercurrent $j$ as a function of $q_{\rm ex}$ corresponds to the SC critical current. In the centrosymmetric case where $q_{\rm ex}$ and $-q_{\rm ex}$ are equivalent to each other, the relation $j(q_{\rm ex})=-j(-q_{\rm ex})$ trivially holds, so that the critical currents in mutually opposite directions, i.e., the positive maximum and negative minimum of $j$, $j_+$ and $j_-$, are equivalent, satisfying $|j_+|=|j_-|$. In the present nonscentrosymmetric case with the RSOC, however, $q_{\rm ex}$ and $-q_{\rm ex}$ are not equivalent and thus, $j(q_{\rm ex})=-j(-q_{\rm ex})$ is not satisfied any more. This means that $|j_+| \neq |j_-|$, namely, the SC critical current becomes nonreciprocal. The ratio 
\begin{equation}
R=\frac{|j_+|-|j_-|}{|j_+|+|j_-|}.
\end{equation} 
measures this nonreciplocity of the critical supercurrent.
 
\begin{figure}[t]
\centering
\includegraphics[width=\columnwidth]{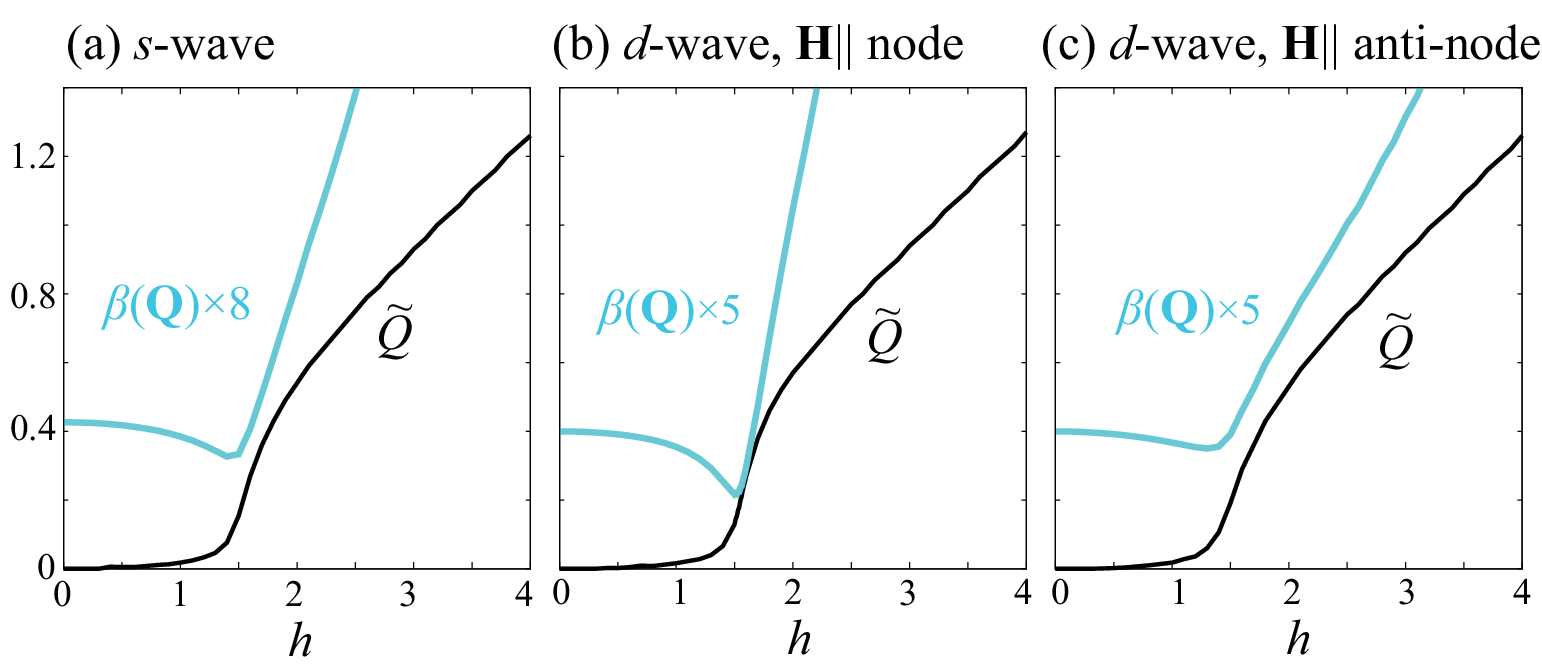}
\caption{Field dependence of the helical modulation $\tilde{Q}$ (black curves) and the coefficient of the $|\Delta_+|^4$ term $\beta({\bf Q})$ (cyan curves) along the $H_{c2}$ curve in the cases of (a) $s$-wave pairing, (b) $d$-wave pairing with ${\bf H}\parallel$ node, and (c) $d$-wave pairing with ${\bf H}\parallel$ anti-node for $\delta N=0.05$. The associated phase diagrams are shown in the lower panels of Figs. \ref{fig:HT} (a), (b), and (c), respectively. \label{fig:Hc2}}
\end{figure}
   
\subsection{Normalization of physical quantities}
In the numerical calculations, we use the following dimensionless parameters 
\begin{equation} 
h= \mu H/T_c, \qquad t=T/T_c. 
\end{equation}
With this normalization, the Pauli limiting field corresponds to $h=1.25$. Also, the SC gap amplitudes $|\Delta_\pm|$, the modulation $Q=|{\bf Q}|$, and the SC current $j$ are normalized as
\begin{equation}
|\tilde{\Delta}_{\pm}|=|\Delta_{\pm}|/T_c, \qquad \tilde{Q}=Q\xi_0, \qquad \tilde{j}=j/j_0, 
\end{equation}
where $\xi_0=v_F/(2\pi T_c)$ is the SC coherence length at $T=0$ and $j_0=2|e|v_F \overline{N}_0 T_c$. In the following results shown in Figs. \ref{fig:Hc2}, \ref{fig:gap_tdep}, and \ref{fig:GL_failure}, the coefficient of the GL quartic term $\beta(\pm{\bf Q})$ is multiplied by $T_c^2$ for nondimensionalization. Concerning the GL quadratic terms $\alpha({\bf Q})$ and $\alpha^{\rm eff}(-{\bf Q})$, they are dimensionless from the beginning.

\section{Stability of the stripe order}
\begin{figure}[t]
\centering
\includegraphics[width=\columnwidth]{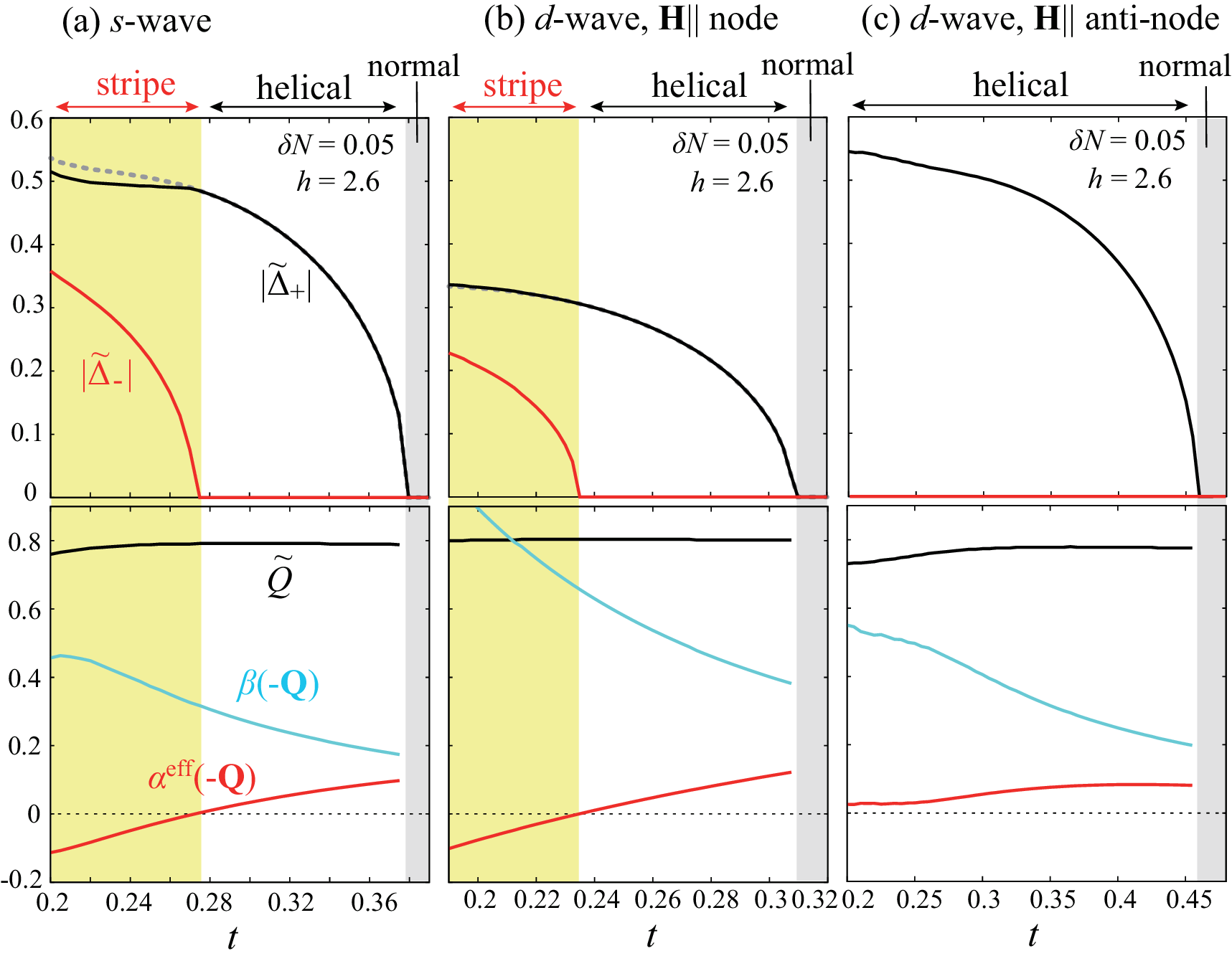}
\caption{Temperature dependence of various physical quantities at $h=2.6$ in the cases of (a) $s$-wave pairing, (b) $d$-wave pairing with ${\bf H}\parallel$ node, and (c) $d$-wave pairing with ${\bf H}\parallel$ anti-node for $\delta N =0.05$. The associated phase diagrams are shown in the lower panels of Fig. \ref{fig:HT}. In the top panels, black and red curves denote the gap amplitudes for the $+{\bf Q}$ and $-{\bf Q}$ modes, $|\tilde{\Delta}_+|$ and $|\tilde{\Delta}_-|$, respectively. For comparison, corresponding results for the helical state are also shown by gray dashed curves. In the bottom panels, black, red, and cyan curves denote the modulation $\tilde{Q}$ and the coefficients of the $|\Delta_-|^2$ and $|\Delta_-|^4$ terms, $\alpha^{\rm eff}(-{\bf Q})$ and $\beta(-{\bf Q})$, respectively. \label{fig:gap_tdep}}
\end{figure}

Figure \ref{fig:HT} shows the temperature and magnetic field phase diagrams in the two-dimensional superconductors with the weak RSOC of $\delta N =0.01$ (upper panels) and the moderate RSOC of $\delta N=0.05$ (lower panels). Although our interest is mainly in the $d$-wave case shown in Figs. \ref{fig:HT} (b) and (c), we show the result for the $s$-wave case in Fig. \ref{fig:HT} (a) for completeness. 
In all the cases shown in Fig. \ref{fig:HT}, the helical SC state with its modulation ${\bf Q}\parallel (\hat{H}\times \hat{z})$ is realized just below the $H_{c2}$ curve. For the $\delta N$ values used here, the direction of ${\bf Q}$ remains unchanged at temperatures lower than but not so far from the $H_{c2}$ transition where the GL theory should work well. Figure \ref{fig:Hc2} shows the field dependence of the helical modulation $\tilde{Q}$ (black curves) and the coefficient of the $|\Delta_+|^4$ term $\beta({\bf Q})$ (cyan curves) along the $H_{c2}$ curve. One can see that around $h=1.6$ slightly above the Pauli limiting field of $h=1.25$, $\tilde{Q}$  exhibits a steep increase and at the same time, $\beta({\bf Q})$ tends to become small. Although the latter tendency is remarkable in the $d$-wave case with ${\bf H} \parallel $ node shown in Fig. \ref{fig:Hc2} (b), $\beta({\bf Q})$ is always positive, so that the transition between the helical SC and normal phases is of second order.  We note in passing that in Fig. \ref{fig:Hc2} (b), due to the sharp drop in $\beta({\bf Q})$ near $h=1.6$, the present GL expansion becomes invalid at further low temperatures (for details, see Appendix A).  

In the reference case of $s$-wave shown in Fig. \ref{fig:HT} (a), one can see that the stripe order is realized in the high-field and low-temperature region. The stripe region for $\delta N=0.05$ is almost quantitatively the same as that in the associated three dimensional system \cite{Stripe_Agterberg_prb_07} in spite of the difference in the dimensionality. The top panels in Fig. \ref{fig:gap_tdep} show the temperature dependence of the gap amplitudes for the $\pm {\bf Q}$ modes $\Delta_{\pm}$ obtained at $h=2.6$ for $\delta N=0.05$. One can see from Fig. \ref{fig:gap_tdep} (a) that with decreasing temperature, $\Delta_+$ first develops and then, $\Delta_-$ starts developing at the helical-stripe transition which is determined by the condition $\alpha^{\rm eff}(-{\bf Q})=0$ and turns out to be of second order as $\beta(-{\bf Q})>0$ (see the lower panel). Since the RSOC prefers a single-${\bf Q}$ helical state, the stripe order involving both ${\bf Q}$ and $-{\bf Q}$ gets unstable with increasing the RSOC [compare the upper and lower panels in Fig. \ref{fig:HT} (a)]. In the associated three dimensional system, it has been reported that for $\delta N=0.25$, the stripe phase cannot exit any more \cite{Stripe_Agterberg_prb_07}.  

In contrast to the $s$-wave case where the SC gap is isotropic and thus, the stability of the stripe order does not depend on the field direction, the stripe ordering in the $d_{x^2-y^2}$ case strongly depends on the field direction due to the existence of the line node running along the [110] and [1$\overline{1}$0] directions. As exemplified by the lower panels of Figs. \ref{fig:gap_tdep} (b) and (c), $\alpha^{\rm eff}(-{\bf Q})$ at the fixed field strength of $h=2.6$ becomes negative at low temperatures for ${\bf H}\parallel $ node, whereas it remains positive for ${\bf H}\parallel$ anti-node, each suggesting the presence and absence of the stripe phase of $\Delta_-\neq 0$. 
The results of this kind at different field strengths are summarized in Fig. \ref{fig:HT}. As readily seen in Figs. \ref{fig:HT} (b) and (c), although the stripe phase can be stable in the high-field and low-temperature region for both the two field configurations, its stability region is significantly suppressed in the ${\bf H}\parallel$ anti-node case. Such a difference between the ${\bf H}\parallel$ node and ${\bf H}\parallel$ anti-node cases becomes more remarkable for stronger RSOC. This field-angle dependence can easily be understood from the FFLO instability in the absence of the RSOC where the LO state with the modulation parallel to the nodal direction is widely stable. Since the RSOC favors the helical modulation perpendicular to the field, the directions of the LO and helical modulations becomes compatible for ${\bf H}\parallel$ node, while not for ${\bf H}\parallel$ anti-node, which results in the angle-dependent stability of the stripe phase. 

In both cases of ${\bf H}\parallel$ node and ${\bf H}\parallel$ anti-node, we have confirmed that the transition between the helical and stripe phases is of second order [for example, see the sign of $\beta(-{\bf Q})$ in Fig. \ref{fig:gap_tdep} (b)], except the low-field phase boundary for ${\bf H}\parallel$ node [see the dashed line in Fig. \ref{fig:HT} (b)]. Near this low-field boundary, the present GL approach taking account of the terms up to the fourth order in the SC gap does not work well (for details, see Appendix A). Nevertheless, considering that in the absence of the RSOC, the low-field boundary is of second order \cite{FFLO_Vorontsov_prb_05}, it is likely to be of second order as well in the presence of the moderate RSOC.

\section{Superconducting diode effect}
In the previous section, we show that the stripe order of the from $\Delta_+ e^{i{\bf Q}\cdot {\bf r}} + \Delta_- e^{-i{\bf Q}\cdot {\bf r}}$ with $|\Delta_+| > |\Delta_-|\neq 0$ can appear via the second order transition from the higher-temperature helical state of the form $\Delta_+ e^{i{\bf Q}\cdot {\bf r}}$. For the $d$-wave pairing, the stripe ordering gets suppressed when the in-plane field is rotated from the nodal direction to the anti-nodal direction. In this section, we will discuss the nonreciprocity of the critical current called the intrinsic SC diode effect \cite{diode_Daido_prl_22, diode_Yuan_pnas_22, diode_He_njp_22, diode_Daido_prb_22}, putting particular emphasis on how the $-{\bf Q}$ mode additionally emerging in the stripe phase affects the nonreciprocity. 
\begin{figure}[t]
\centering
\includegraphics[width=\columnwidth]{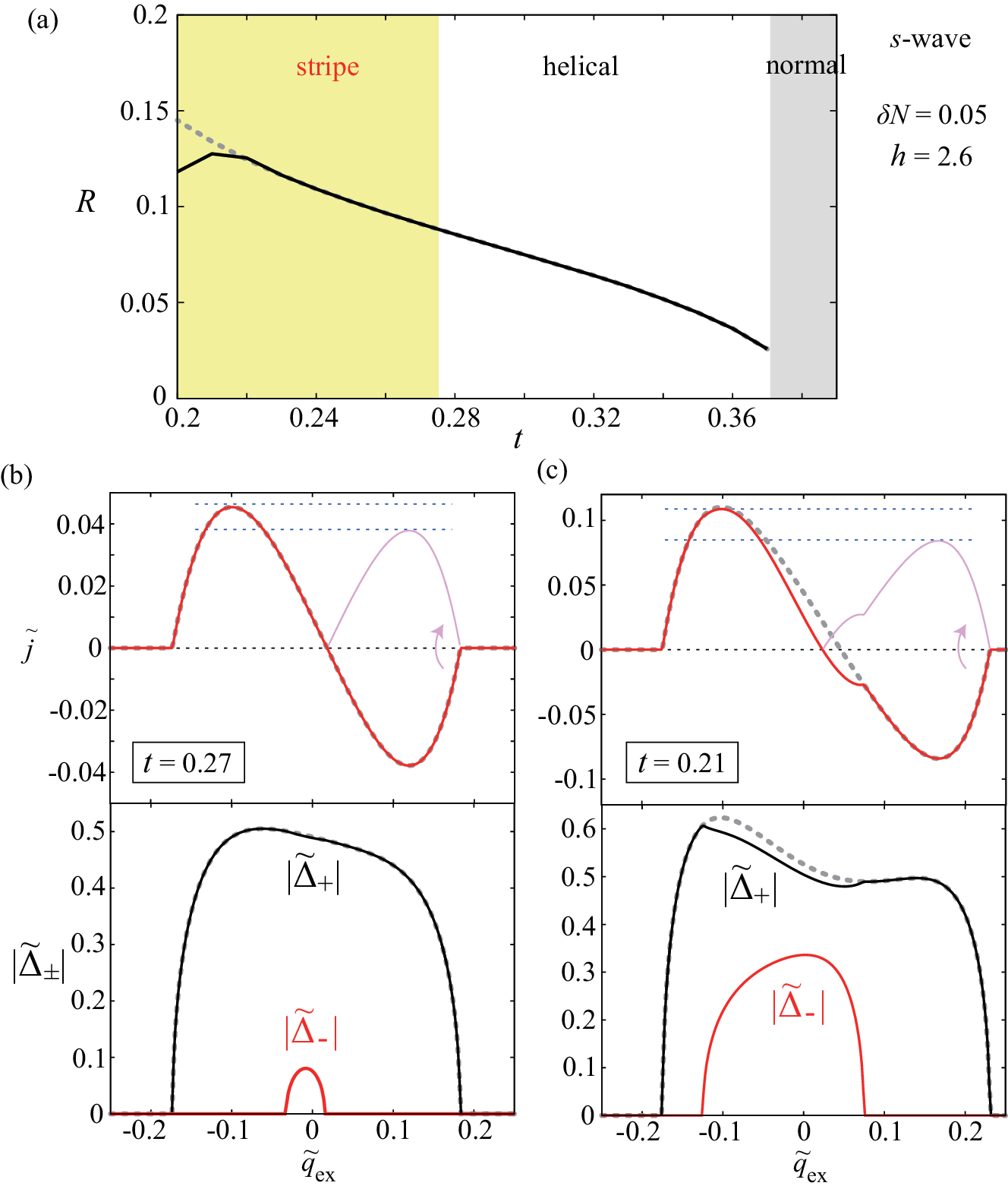}
\caption{The nonreciprocity of the critical current obtained at $h=2.6$ in the $s$-wave case with $\delta N=0.05$, where the system parameters are the same as those in Fig. \ref{fig:gap_tdep} (a). (a) The temperature dependence of the nonreciprocity $R$ and (b) [(c)] the $\tilde{q}_{\rm ex}$ dependence of the current $\tilde{j}$ (upper panel)  and the gap amplitudes $|\tilde{\Delta}_+|$ and $|\tilde{\Delta}_-|$ (black and red curves in the lower panel) at $t=0.27$ ($t=0.21$). For comparison, corresponding results for the helical state are also shown by gray dashed curves. In the upper panels of (b) and (c), the negative part of $\tilde{j}$ is folded back (see pink curves), and the peak-height difference is indicated by blue dotted lines. \label{fig:diode_s}}
\end{figure}

We shall start from the $s$-wave case. Figure \ref{fig:diode_s} (a) shows the temperature dependence of the nonreciprocity $R$ in the $s$-wave case with $\delta N=0.05$ and $h=2.6$, the same parameter set as that in Fig. \ref{fig:gap_tdep} (a). In Fig. \ref{fig:diode_s} (a), in addition to the main result represented by the solid black curve, we show, for comparison, the result obtained under the constraint of $\Delta_-=0$ for which only the helical state is allowed (the gray dashed curve). One can see that $R$ is nonzero not only in the helical phase but also in the stripe phase and that a signature of the second order transition into the stripe phase cannot be found in $R$. At further low temperatures, the nonreciprocity $R$ in the stripe phase becomes slightly smaller than the one for the helical state [compare the black solid and gray dashed curves in Fig. \ref{fig:diode_s} (a)]. To understand these behaviors of $R$, we shall look into the details of the supercurrent $\tilde{j}$ as a function of $\tilde{q}_{\rm ex}$ corresponding to the external current.

\begin{figure}[t]
\centering
\includegraphics[width=\columnwidth]{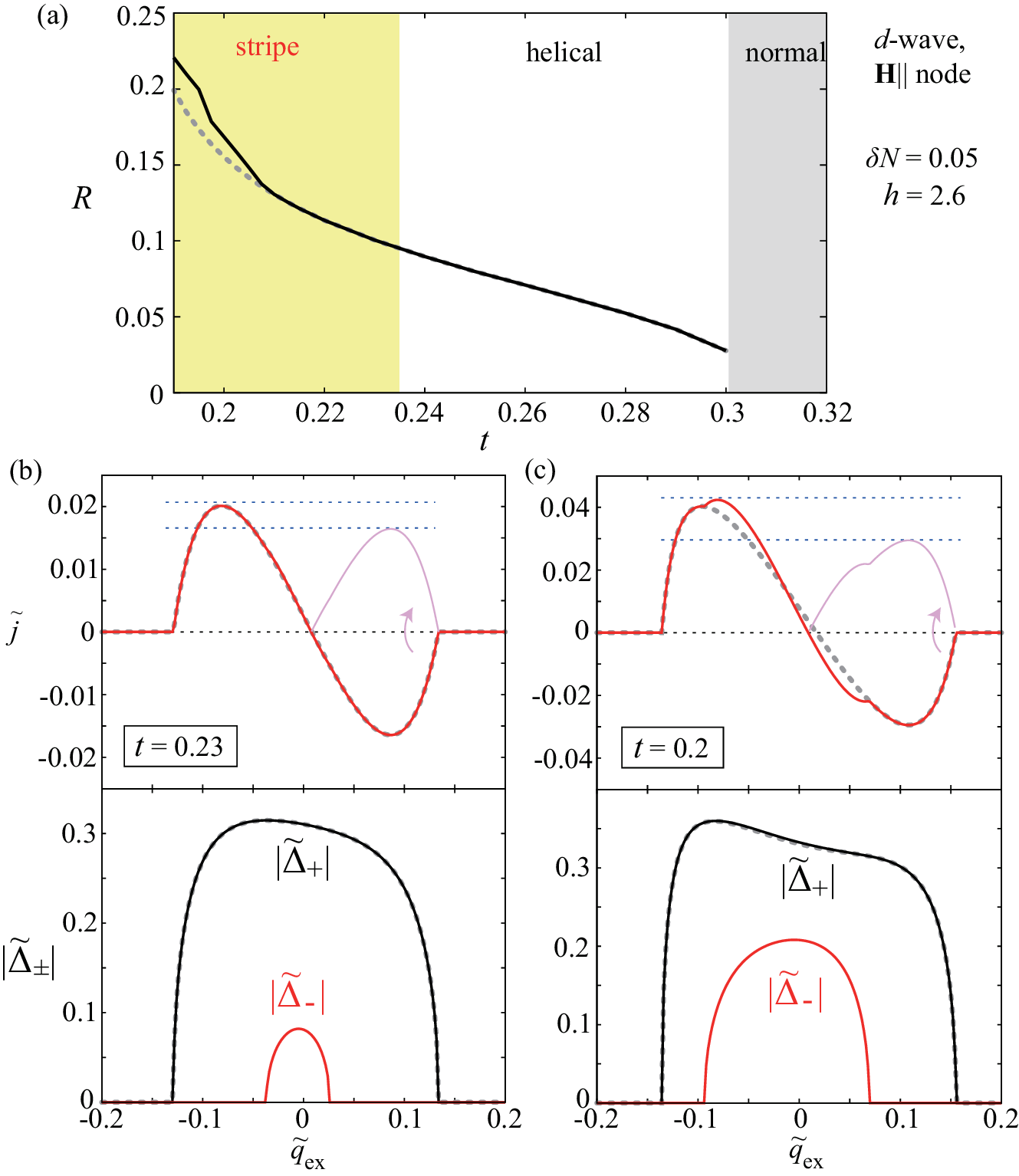}
\caption{The critical-current nonreciprocity obtained at $h=2.6$ in the $d$-wave case with ${\bf H}\parallel$ node and $\delta N=0.05$, where the system parameters are the same as those in Fig. \ref{fig:gap_tdep} (b). (a) The temperature dependence of $R$ and (b) [(c)] the $\tilde{q}_{\rm ex}$ dependence of $\tilde{j}$ and $|\tilde{\Delta}_{\pm}|$ at $t=0.23$ ($t=0.2$). The color and symbol notations are the same as those in Fig. \ref{fig:diode_s}.  \label{fig:diode_d-node}}
\end{figure}

Figures \ref{fig:diode_s} (b) and (c) show the $\tilde{q}_{\rm ex}$ dependence of $\tilde{j}$ (upper panels) and the gap amplitudes $|\tilde{\Delta}_\pm|$ (lower panels) at $t=0.27$ and $t=0.21$, respectively, where in the upper panels, the negative part of $\tilde{j}$ is folded back (see the pink curve) such that the nonreciprocity, which corresponds to the peak-height difference indicated by blue dotted lines, can easily be confirmed. In Figs. \ref{fig:diode_s} (b) and (c), the notation of the gray dashed curve is the same as that in Fig. \ref{fig:diode_s} (a); it represents the result obtained by assuming that only the helical phase is realized. One can see from Fig. \ref{fig:diode_s} (b) that just below the helical-stripe transition, the $-{\bf Q}$ component $\tilde{\Delta}_-$, which is very small due to the second-order nature of the transition, is rapidly suppressed by $\tilde{q}_{\rm ex}$ (see the lower panel) and does not affect the peak heights of $\tilde{j}$, or equivalently, the critical currents (see the upper panel). Therefore, the nonreciprocity $R$ in the stripe phase just below the transition is the same as that for the helical phase, and does not show any clear anomaly at the transition. Further below the transtion, on the other hand, as shown in Fig. \ref{fig:diode_s} (c), $|\tilde{\Delta}_-|$ is relatively robust against $\tilde{q}_{\rm ex}$, so that it can affect the critical current. The peak hight of the positive part of $\tilde{j}$ is slightly suppressed in the stripe phase [compare the red solid and gray dashed curves in Fig. \ref{fig:diode_s} (c)], and correspondingly, the nonreciprocity $R$ takes a smaller value [see Fig. \ref{fig:diode_s}(a)]. 

\begin{figure}[t]
\centering
\includegraphics[width=\columnwidth]{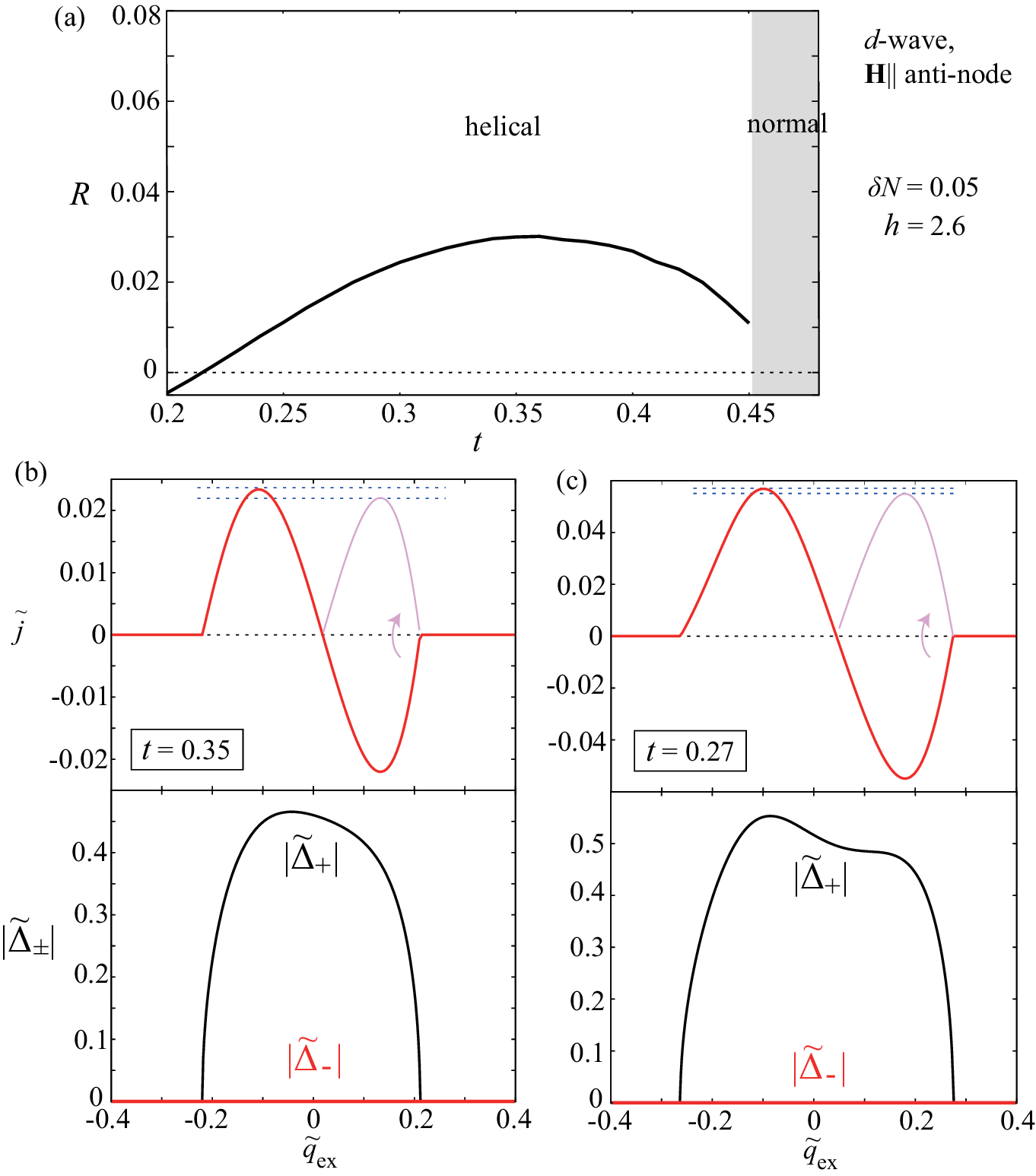}
\caption{The critical-current nonreciprocity obtained at $h=2.6$ in the $d$-wave case with ${\bf H}\parallel$ anti-node and $\delta N=0.05$, where the system parameters are the same as those in Fig. \ref{fig:gap_tdep} (c). (a) The temperature dependence of $R$ and (b) [(c)] the $\tilde{q}_{\rm ex}$ dependence of $\tilde{j}$ and $|\tilde{\Delta}_{\pm}|$ at $t=0.35$ ($t=0.27$). The color and symbol notations are the same as those in Fig. \ref{fig:diode_s}. \label{fig:diode_d-antinode}}
\end{figure}

\begin{figure*}[t]
\centering
\includegraphics[width=2\columnwidth]{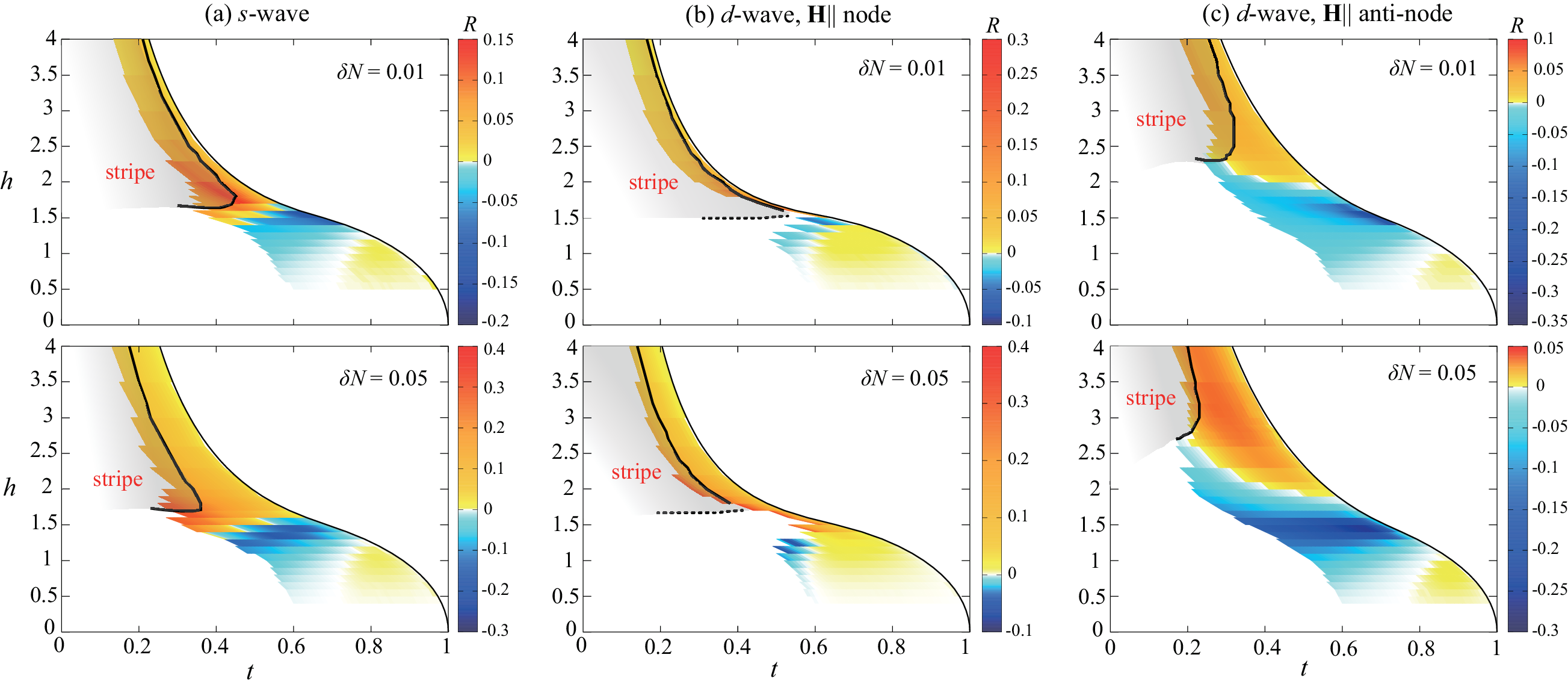}
\caption{Temperature and magnetic-field dependence of the nonreciprocity $R$ in the cases of (a) $s$-wave pairing, (b) $d$-wave pairing with ${\bf H}\parallel$ node, and (c) $d$-wave pairing with ${\bf H}\parallel$ anti-node for $\delta N=0.01$ (upper panels) and $\delta N=0.05$ (lower panels), where reddish and bluish colors represent positive and negative values of $R$, respectively. The stripe region in Fig. \ref{fig:HT} is indicated by translucent gray in this figures. \label{fig:diode_HT}}
\end{figure*}

We next discuss the $d$-wave case. Figure \ref{fig:diode_d-node} shows the result on the nonreciprocity in the $d$-wave case with ${\bf H}\parallel $ node, where the notations are the same as those in the $s$-wave case of Fig. \ref{fig:diode_s}. One can see from Fig. \ref{fig:diode_d-node} (a) that the critical-current nonreciprocity emerges in the stripe phase without showing an anomaly at the helical-stripe transition. The reason for the absence of the signature of the transition is the same as that in the $s$-wave case: just below the transition, the $-{\bf Q}$ component $\tilde{\Delta}_-$ is fragile against the external current $\tilde{q}_{\rm ex}$ [see Fig. \ref{fig:diode_d-node} (b)]. At further low temperatures, there is a difference between the $s$-wave and $d$-wave cases. In Fig. \ref{fig:diode_d-node} (c), although it is common that $\tilde{\Delta}_-$ is relatively robust against $\tilde{q}_{\rm ex}$, the positive peak of $\tilde{j}$ is not suppressed but rather enhanced by the contribution from $\tilde{\Delta}_-$, which results in a slight increase in $R$ [compare the black solid and gray dashed curves in Fig. \ref{fig:diode_d-node} (a)]. In experiments, however, such a slight deviation from the helical value could not be captured, since 
as will be discussed below, even in the helical phase, the temperature dependence of $R$ is not so simple.   

Figure \ref{fig:diode_d-antinode} (a) shows the temperature dependence of $R$ for ${\bf H}\parallel$ anti-node, where only the helical phase is realized over the temperature range of this figure. The nonreciprocity $R$ exhibits a non-monotonic temperature dependence. Note that $\tilde{\Delta}_-$ remains zero even after $\tilde{q}_{\rm ex}$ is introduced [see Figs. \ref{fig:diode_d-antinode} (b) and (c)] and thus, this behavior is purely of helical origin. In addition, the value of $R$ for ${\bf H}\parallel$ anti-node is much smaller than $R$ for ${\bf H}\parallel$ node and takes negative sign at the lowest temperature, reflecting the fact that, as will be explained below, $R$ changes its sign at lower fields.

Figure \ref{fig:diode_HT} shows the temperature and magnetic-field dependence of $R$ for the same parameter set as that for Fig. \ref{fig:HT}, where the reddish and bluish colors represent positive and negative values of $R$, respectively, and the stability region of the stripe phase in Fig. \ref{fig:HT} is indicated by translucent gray. In Fig. \ref{fig:diode_HT} (b), data blanks near $h=1.6$ are due to the invalidity of the GL approach used here (see Sec. II and Appendix A). As already reported elsewhere \cite{diode_Daido_prl_22, diode_Daido_prb_22}, the nonreciprocity $R$ tends to changes its sign near the field (in the present case, $h \sim 1.6$) at which the helical modulation $Q$ rapidly develops (see Fig. \ref{fig:Hc2}). In the $d$-wave case with the moderate RSOC of $\delta N=0.05$, however, the sign change is suppressed when the magnetic field is applied along the nodal direction [compare the lower panels of Figs. \ref{fig:diode_HT} (b) and (c)]. Although the sign of $R$ seems to depend on the details of specific systems such as the Fermi surface shape controlled by electron fillings \cite{diode_Daido_prl_22, diode_Daido_prb_22}, the above result obtained for the isotropic cylindrical Fermi surface should capture the essential part of roles of the gap anisotropy. Concerning the main focus of this work, i.e., how the nonreciprocity looks like in the stripe phase, it continuously changes on cooling across the transition from the helical phase into the stripe phase, and any characteristic feature such as a sign change in $R$ cannot be found at the transition, being irrespective of the orbital pairing symmetry, the strength of the RSOC, and the field direction.

\section{Summary and discussion}
In this work, we have theoretically investigated the stability of the stripe order, a LO-like state described as a superposition of the $+{\bf Q}$ and $-{\bf Q}$ modes of different weight, in two-dimensional superconductors with the Rashba spin-obrit coupling (RSOC) and an in-plane magnetic field where the helical SC state with only the $+{\bf Q}$ mode is widely stabilized. Based on the GL analysis, we show that the stripe phase can be stabilized in the high-field and low-temperature region for both the $s$-wave and $d$-wave pairing symmetries, as originally pointed out for the three-dimensional $s$-wave case \cite{Stripe_Agterberg_prb_07}. Interestingly, in the $d$-wave case, the stability region of the stripe phase shrinks when the in-plane field is merely rotated from the nodal direction to the anti-nodal direction. It is also found that the nonreciprocity of the critical current, the so-called the SC diode effect, emerges not only in the helical phase but also in the stripe phase. The transition between the helical and stripe phases does not leave a footprint in the temperature dependence of the critical-current nonreciprocity due to the second-order nature of the transition.

In experiments on film superconductors, it is usually difficult to perform bulk measurements such as the specific heat, so that transport measurements have widely been used to study SC states. Although a signature of the stripe phase has not been observed so far in relevant two-dimensional Rashba systems \cite{diode_Ando_nature_20, superlattice_Naritsuka_prb_17, interface-SC, tunable-SC, tunable-RSO, gateE-SC_Ueno_natmat_08, gateE-SC_Ueno_prb_14}, this might be simply because the RSOC is too strong for the stripe order to survive or because as discussed above, the critical current is insensitive to the second order transition into the stripe phase. Also, a current-driven dynamics of vortices could contribute to the SC transport. First, in two dimensions, thermally activated Kosterlitz-Thouless vortices with their axis parallel to the out-of-plane direction can appear, but they are active basically at higher temperatures near the SC transition \cite{diode_Hoshino_prb_18} and thus, should be irrelevant to the low-temperature transport. Second, in real experimental systems with finite thickness in the out-of-plane direction, the in-plane field may yield SC vortices extending along the in-plane field direction and a current-driven out-of-plane motion of these vortices could be relevant. 
If the vortex motion is affected by the stripe modulation perpendicular to the vortex line, we may have a chance to detect a signature of the stripe order in the vortex dynamics.

Even in the presence of the vortex dynamics, the nonreciprocity of the critical current should more or less exist. In contrast to the isotropic $s$-wave case, in the anisotropic $d$-wave case, the sign of the nonreciprocity is dependent on the in-plane field direction  relative to the nodal direction (see Fig. \ref{fig:diode_HT}). A recent transport measurement on the tricolor superlattice of YbRhIn$_5$/CeCoIn$_5$/YbCoIn$_5$, the two-dimensional Rashba surperconductor with the $d_{x^2-y^2}$ pairing symmetry, has shown that the field dependence of the nonreciprocity exhibits an anomaly for ${\bf H}\parallel$ [100], while not for ${\bf H}\parallel$ [110] \cite{superlattice_Matsuda_private}, which might be a manifestation of the field-angle-dependent sign change in the critical-current nonreciprocity originating from the $d$-wave anisotropy. 
Although the Fermi surface shape, which is assumed, for simplicity, to be cylindrical in this work, could quantitatively affect the results and thus, should carefully be considered in our future work, we believe that our result obtained for the simplified model captures the essential part of the stripe ordering and the associated nonreciprocal phenomena in two-dimensional Rashba superconductors. 

\begin{acknowledgements}
We are grateful to Y. Matsuda, Y. Kasahara, and A. Daido for useful discussions. This work is partially supported by JSPS KAKENHI Grant No. JP21K03469 and JP23H00257.
\end{acknowledgements}

\appendix
\section{Applicability range of the GL expansion}
\begin{figure}[b]
\centering
\includegraphics[width=\columnwidth]{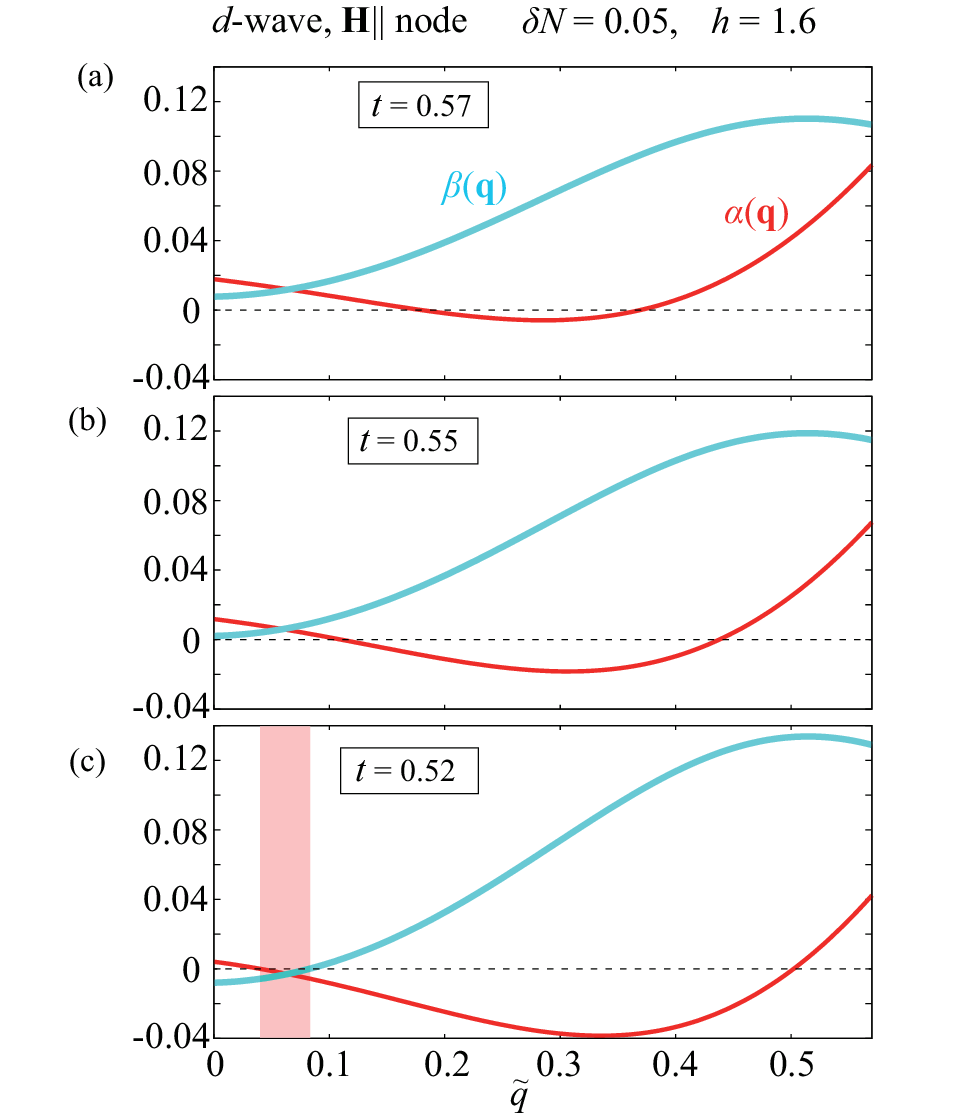}
\caption{The GL coefficients for the helical state, $\alpha({\bf q})$ (red curves) and $\beta({\bf q})$ (cyan curves), as a function of the center-of-mass momentum of the Cooper pair  $\tilde{q}$ at (a) $t=0.57$, (b) $t=0.55$, and (c) $t=0.52$ in the $d$-wave case with $\delta N=0.05$ and $h=1.6$, where the in-plane field is applied in the nodal direction. \label{fig:GL_failure}}
\end{figure}

In general, the GL expansion up to the fourth order is justified when the coefficient of the quartic term is positive. 
As one can see from Fig. \ref{fig:Hc2} (b), the coefficient of the $|\Delta_+|^4$ term $\beta({\bf Q})$ at the SC transition temperature defined by $\alpha({\bf Q})=0$ tends to approach zero near $h=1.6$, suggesting that the GL expansion may possibly become invalid around this field and temperature. Figure \ref{fig:GL_failure} shows the $q(=|{\bf q}|)$ dependence of $\alpha({\bf q})$ and $\beta({\bf q})$ at various temperatures for $h=1.6$. One can see from Fig. \ref{fig:GL_failure} (a) that at $t=0.57$ just below the $H_{c2}$ transition, $\beta({\bf q})$ is always positive for $q$ satisfying $\alpha({\bf q})<0$, so that the optimal modulation $q=Q$ minimizing the condensation energy can definitely be identified. With decreasing temperature,  $\beta({\bf q})$ near $q\sim 0$ gradually decreases, eventually taking a negative sign at $t=0.52$. In Fig. \ref{fig:GL_failure} (c), one notices that $\beta({\bf q})$ can be negative in the ordered state of $\alpha({\bf q})<0$ (see the colored window), so that neither the optimal modulation nor the gap amplitude can be determined. Due to the invalidity of the GL expansion, we cannot discuss the SC properties near $h=1.6$ in the $d$-wave case with ${\bf H}\parallel $ node, and thus, we draw a putative low-field boundary of the stripe phase with a dashed line in Fig. \ref{fig:HT} (b) and leave the invalid region blank in Fig. \ref{fig:diode_HT} (b).  

Inside the stripe phase where we have the two components $\Delta_+$ and $\Delta_-$, a condition to guarantee the validity of the GL expansion is $D>0$, where $D$ is defined in Eq. (\ref{eq:GL_sol}). Note that if $D<0$ in Eq. (\ref{eq:GL_sol}), $|\Delta_+|^2$ and/or $|\Delta_-|^2$ can become negative, namely, the solution of the GL equations can be unphysical. In the terms of the original GL free energy (\ref{eq:fGL_stripe}), the negative $D$ means that when the quartic terms in $\Delta_+$ and/or $\Delta_-$ are diagonalized, one eigen value becomes negative, i.e., one of the diagonalized quartic terms has a negative coefficient, and thus, a local minimum does not exit. In this paper, we restrict ourselves to the parameter space where $D>0$ is satisfied, and the results obtained within the parameter space are shown. 
In principle, the above problems could be resolved by taking higher order contributions into account, but this issue is beyond the scope of this work.


\begin{thebibliography}{50}
\bibitem{FF} P. Fulde and R. A. Ferrell, Phys. Rev. {\bf 135}, A550 (1964). 
\bibitem{LO} A. I. Larkin and Yu. N. Ovchinnikov, Sov. Phys. JETP {\bf 20}, 762 (1965). 
\bibitem{FFLO_Matsuda_review_07} Y. Matsuda and H. Shimahara, J. Phys. Soc. Jpn. {\bf 76}, 051005 (2007).

\bibitem{Vorontsov_SF} A. B. Vorontsov and J. A. Sauls, Phys. Rev. Lett. {\bf 98}, 045301 (2007).
\bibitem{Vorontsov_SC} A. B. Vorontsov, Phys. Rev. Lett. {\bf 102}, 177001 (2009).
\bibitem{Hachiya} M. Hachiya, K. Aoyama, and R. Ikeda, Phys. Rev. B {\bf 88}, 064519 (2013).
\bibitem{Aoyama_cylinder} K. Aoyama, Phys. Rev. B {\bf 89}, 140502(R) (2014).
\bibitem{Aoyama_film} K. Aoyama, J. Phys. Soc. Jpn. {\bf 85}, 094604 (2016).
\bibitem{Wiman_film} J. J. Wiman and J. A. Sauls, Phys. Rev. B {\bf 92}, 144515 (2015).

\bibitem{Stripe_Agterberg_prb_07} D. F. Agterberg and R. P. Kaur, Phys. Rev. B {\bf 75}, 064511 (2007).
\bibitem{2layer-Stripe_Yoshida_jpsj_13} T. Yoshida, M. Sigrist, and Y. Yanase, J. Phys. Soc. Jpn. {\bf 82}, 074714 (2013).

\bibitem{Kaur} R. P. Kaur, D. F. Agterberg, and M. Sigrist, Phys. Rev. Lett. {\bf 94}, 137002 (2005).
\bibitem{Springer} {\it Non-Centrosymmetric Superconductors: Introduction and Overview (Lecture Notes in Physics)}, edited by E. Bauer and M. Sigrist, Springer 2012.

\bibitem{diode_Wakatsuki_prl_18} R. Wakatsuki and N. Nagaosa, Phys. Rev. Lett. {\bf 121}, 026601 (2018). 
\bibitem{diode_Wakatsuki_sadv_17} R.Wakatsuki, Y. Saito, S. Hoshino, Y. M. Itahashi, T. Ideue, M. Ezawa, Y. Iwasa, and N. Nagaosa, Sci. Adv. {\bf 3}, e1602390 (2017). 

\bibitem{diode_Ando_nature_20} F. Ando, Y. Miyasaka, T. Li, J. Ishizuka, T. Arakawa, Y. Shiota, T. Moriyama, Y. Yanase, and T. Ono, Nature(London) {\bf 584}, 373 (2020). 
\bibitem{diode_Daido_prl_22} A. Daido, Y. Ikeda, and Y. Yanase, Phys. Rev. Lett. {\bf 128}, 037001 (2022). 
\bibitem{diode_Yuan_pnas_22} N. F. Q. Yuan and L. Fu, Proc. Natl. Acad. Sci. U. S. A. {\bf 119}, e2119548119 (2022). 
\bibitem{diode_He_njp_22} J. J. He, Y. Tanaka, N. Nagaosa, New. J. Phys. {\bf 24}, 053014 (2022). 
\bibitem{diode_Daido_prb_22} A. Daido and Y. Yanase, Phys. Rev. B {\bf 106}, 205206 (2022). 

\bibitem{CePt3Si} E. Bauer, G. Hilscher, H. Michor, Ch. Paul, E. W. Scheidt, A. Gribanov, Yu. Seropegin, H. Noel, M. Sigrist, and P. Rogl, Phys. Rev. Lett. {\bf 92} 027003 (2004). 
\bibitem{CeRhSi3} N. Kimura, K. Ito, K. Saitoh, Y. Umeda, H. Aoki, and T. Terashima, Phys. Rev. Lett. {\bf 95} 247004 (2005). 
\bibitem{CeIrSi3} I. Sugitani, Y. Okuda, H. Shishido, T. Yamada, A. Thamizhavel, E. Yamamoto, T. D. Matsuda, Y. Haga, T. Takeuchi, R. Settai, and Y. Onuki, J. Phys. Soc. Jpn. {\bf 75} 043703 (2006). 
\bibitem{superlattice_Goh_prl_12}  S. K. Goh, Y. Mizukami, H. Shishido, D. Watanabe, S. Yasumoto, M. Shimozawa, M. Yamashita, T. Terashima, Y. Yanase, T. Shibauchi, A. I. Buzdin, and Y. Matsuda, Phys. Rev. Lett. {\bf 109}, 157006 (2012).
\bibitem{superlattice_Shimozawa_prl_14} M. Shimozawa, S. K. Goh, R. Endo, R. Kobayashi, T. Watashige, Y. Mizukami, H. Ikeda, H. Shishido, Y. Yanase, T. Terashima, T. Shibauchi, and Y. Matsuda, Phys. Rev. Lett. {\bf 112}, 156404 (2014).
\bibitem{superlattice_Naritsuka_prb_17} M. Naritsuka, T. Ishii, S. Miyake, Y. Tokiwa, R. Toda, M. Shimozawa, T. Terashima, T. Shibauchi, Y. Matsuda, and Y. Kasahara, Phys. Rev. B {\bf 96}, 174512 (2017).
\bibitem{interface-SC} N. Reyren, S. Thiel, A. D. Caviglia, L. Fitting Kourkoutis, G. Hammerl, C. Richter, C. W. Schneider, T. Kopp, A. S. Retschi, D. Jaccard, M. Gabay, D. A. Muller, J.-M. Triscone, and J. Mannhart, Science {\bf 317}, 1196 (2007).
\bibitem{tunable-SC} A. D. Caviglia, S. Gariglio, N. Reyren, D. Jaccard, T. Schneider, M. Gabay, S. Thiel, G. Hammerl, J. Mannhart, and J.-M. Triscone, Nature(London) {\bf 456}, 624 (2008).
\bibitem{tunable-RSO} A. D. Caviglia, M. Gabay, S. Gariglio, N. Reyren, C. Cancellieri, and J. -M. Triscone, Phys. Rev. Lett. {\bf 104}, 126803 (2010).
\bibitem{gateE-SC_Ueno_natmat_08} K. Ueno, S. Nakamura, H. Shimotani, A. Ohtomo, N. Kimura, T. Nojima, H. Aoki, Y. Iwasa, and M. Kawasaki, Nat. Mater. {\bf 7}, 855 (2008). 
\bibitem{gateE-SC_Ueno_prb_14} K. Ueno, T. Nojima, S. Yonezawa, M. Kawasaki, Y. Iwasa, and Y. Maeno, Phys. Rev. B {\bf 89}, 020508(R) (2014).


\bibitem{SC_jM} V. M. Edelstein, Phys. Rev. Lett. {\bf 75}, 2004 (1995). 
\bibitem{SC_Bj} V. M. Edelstein, Sov. Phys. JETP {\bf 68}, 1244 (1989); S. K. Yip, Phys. Rev. B {\bf 65}, 144508 (2002). 
\bibitem{Dimitrova} O. V. Dimitrova and M. V. Feigel'man, JETP Lett. {\bf 78}, 637 (2003).
\bibitem{Samokhin} K. V. Samokhin, Phys. Rev. B {\bf 70}, 104521 (2004).
\bibitem{Fujimoto} S. Fujimoto, Phys. Rev. B {\bf 72}, 024515 (2005). 
\bibitem{Helical_AS_12} K. Aoyama and M. Sigrist, Phys. Rev. Lett. {\bf 109} 237007 (2012).
\bibitem{Helical_ASS_14} K. Aoyama, L. Savary, and M. Sigrist, Phys. Rev. B {\bf 89}, 174518 (2014).

\bibitem{Ikeda_0} Y. Matsunaga, N. Hiasa, and R. Ikeda, Phys. Rev. B {\bf 78}, 220508(R) (2008).

\bibitem{FFLO_d_Maki_physicaB_02} K. Maki and H. Won, Physica B {\bf 322}, 315 (2002).
\bibitem{FFLO_Vorontsov_prb_05} A. B. Vorontsov, J. A. Sauls, and M. J. Graf, Phys. Rev. B {\bf 72}, 184501 (2005).  

\bibitem{VLFF} A. D. Bianchi, M. Kenzelmann, L. DeBeer-Schmitt, J. S. White, E. M. Forgan, J. Mesot, M. Zolliker, J. Kohlbrecher, R. Movshovich, E. D. Bauer, J. L. Sarrao, Z. Fisk, C. Petrovic, and M. R. Eskildsen, Science {\bf 319}, 177 (2008).  
\bibitem{Hiasa} N. Hiasa and R. Ikeda, Phys. Rev. Lett. {\bf 101}, 027001 (2008). 
\bibitem{Magreso_Stock_prl_08} C. Stock, C. Broholm, J. Hudis, H. J. Kang, and C. Petrovic, Phys. Rev. Lett. {\bf 100}, 087001 (2008).
\bibitem{angledep_Vorontsov} A. Vorontsov and I. Vekhter, Phys. Rev. Lett. {\bf 96}, 237001 (2006); Phys. Rev. B {\bf 75}, 224501 (2007). 
\bibitem{d-wave} K. An, T. Sakakibara, R. Settai, Y. Onuki, M. Hiragi, M. Ichioka, and K. Machida, Phys. Rev. Lett. {\bf 104}, 037002 (2010). 
\bibitem{node-imaging_Allan_natphys_13} M. P. Allan, F. Massee, D. K. Morr, J. Van Dyke, A. W. Rost, A. P. Mackenzie, C. Petrovic, and J. C. Davis, Nat. Phys. {\bf 9}, 468 (2013).
\bibitem{QPI_Zhou_natphys_13} B. B. Zhou, S. Misra, E. H. da Silva Neto, P. Aynajian, R. E. Baumbach, J. D. Thompson, E. D. Bauer, and A. Yazdani, Nat. Phys. {\bf 9}, 474 (2013).

\bibitem{superlattice_Matsuda_private} Y. Matsuda and Y. Kasahara, private communication.


\bibitem{Ikeda} N. Hiasa, T. Saiki, and R. Ikeda, Phys. Rev. B {\bf 80}, 014501 (2009). 
\bibitem{AFMSC_KA_prb_11} K. Aoyama and R. Ikeda, Phys. Rev. B {\bf 84}, 184516 (2011). 



\bibitem{Stripe_Dimitrova_prb_07} Ol'ga Dimitrova and M. V. Feigel'man, Phys. Rev. B {\bf 76}, 014522 (2007).

\bibitem{Thinkham} M. Thinkham, {\it Introduction to Superconductivity} (Dover, New York, 1996) Second edition, chap.4. 
\bibitem{LP_KA_prb} K. Aoyama, Phys. Rev. B {\bf 106}, L060502 (2022); Phys. Rev. B {\bf 108}, L060502 (2023). 

\bibitem{diode_Hoshino_prb_18} S. Hoshino, R. Wakatsuki, K. Hamamoto, and N. Nagaosa, Phys. Rev. B {\bf 98}, 054510 (2018). 

\end{thebibliography}
\end{document}